\documentclass[aps,prl,reprint,showpacs,preprintnumbers,amsmath,amssymb]{revtex4-1}

\usepackage{graphicx}
\usepackage{dcolumn}
\usepackage{bm}
\usepackage{color}

\usepackage{amsmath}
\usepackage{amsfonts}
\usepackage{mathtools}
\allowdisplaybreaks

\newcommand{\mi}{\mathrm{i}}
\begin{document}
\def\v#1{{\bf #1}}

\title{Gapless Triplet Superconductivity in Magnetically Polarized Media}

\author{Marios Georgiou}%\email{Electronic address: mgeor@mail.ntua.gr}
\author{Georgios Varelogiannis}
\affiliation{Department of Physics, National Technical University
of Athens, GR-15780 Athens, Greece}

\vskip 1cm
\begin{abstract}

We reveal that
in a magnetically polarized medium, a specific 
triplet commensurate pair density wave superconducting (SC) state, the staggered
d-wave $\Pi$-triplet state,
may coexist with homogeneous triplet SC states and even
dominate eliminating them under generic conditions.
When only this TPDW SC state is present,
we have the remarkable
phenomenon of \emph{gapless superconductivity}.
This may explain part of the difficulties in the realization of
the engineered localized Majorana fermion modes for topological quantum computation.
We point out qualitative characteristics of the
tunneling density of states, specific heat and charge susceptibility
that identify the accessible triplet SC regimes in a spinless medium.

\end{abstract}

\pacs{74.81.-g, 74.20.Rp, 74.25.Dw}

\maketitle

Singlet superconductivity (SC) and ferromagnetism (FM) are directly competing phenomena.
The discovery of SC coexisting with FM in UGe$_2$ \cite{Saxena},
and other bulk FM-SC \cite{Aoki, Huy},
in heterostructures where proximity of SC and FM is enforced \cite{Kaizer,Wang,Khaire}
necessarily involves
exotic spin-triplet SC states.
Numerous theoretical models
with homogeneous triplet SC states possibly odd in frequency
have been proposed \cite{Efetov,Buzdin,Eschrig,Eschrig1,Volkov1}.
For the 2-D SC state that develops at the interfaces
of some oxide insulators like LaAlO$_3$/SrTiO$_3$
\cite{InterfaceInsSC} in the presence of FM \cite{FMSCinInterfaces} a
modulated or Pair Density Wave (PDW) triplet state of FFLO type
has also been suggested \cite{PALeePDWinterfaces}.

The observation of proximity induced SC in the half
metallic (fully polarized) FM CrO$_2$
in contact with SC NbTiN
\cite{Kaizer} demonstrates that effectively spinless systems
may exhibit SC as well.
Triplet SC in spinless systems is of
enormous interest
because a spinless triplet SC wire can exhibit
at its two edges localized Majorana fermion modes \cite{Majorana,TopologicQcomp}
Such localized
Majorana fermions \cite{Majorana}
would allow for non-local quantum information storage avoiding local decoherence as
well as for logical manipulations through braiding
because of their non-abelian character \cite{TopologicQcomp}.
The \emph{localization} of these modes
is a crucial requirement
for quantum-bit realizations
and braiding manipulations, and it can occur only if a \emph{finite SC gap}
is present\cite{GapLocalization}.

In the present Letter, based on a systematic study of the interplay of all SC condensates
allowed by symmetry in a fully spin-polarized medium,
we show that under realistic generic conditions
a {\it triplet} SC state exhibiting a \emph{commensurate} density wave modulation
of the superfluid density may coexist with, or even dominate eliminating it,
the homogeneous (zero momentum) triplet SC.
When this {\it triplet commensurate pair density wave SC (TCPDWSC)}
state dominates we have robust
\emph{gapless} SC,
a situation that would be catastrophic for the engineered
topological quantum bits.
We report phase transitions between the various types of accessible
triplet SC states including \emph{transitions between gapped and gapless SC states},
as well as qualitative physical characteristics in the density of states,
specific heat and charge susceptibility
that would allow to identify the type of triplet SC in which the
system of interest is in.

A similar TCPDWSC state channel has
been suggested to occur in the high field SC state of
CeCoIn$_{5}$ coexisting with singlet SC and spin density waves \cite{Aperis,Yanase} explaining
fascinating neutron scattering results \cite{Kenzelmann}.
There have also been studies of TCPDWSC in the singlet channel,
also called $\eta$-pairing, \cite{Loder,Fradkin1, Lee,Fradkin2, Chubukov,Corboz}
mainly motivated by
the extraordinary physics in the pseudogap and other stripe regimes of cuprates
\cite{Zhang,Berg} where extended FS in the SC state has been reported as well \cite{Fradkin1}.

Our starting point is a BCS-type Hamiltonian with frozen spin:
%\begin{align}
%\label{hamiltonian}
$\mathcal{H} = \sum_{\mathbf{k}}\xi_{\mathbf{k}}\,c^{\dagger}_{\mathbf{k}}c_{\mathbf{k}}-
\sum_{\mathbf k} ( \Delta^{\mathbf{0}}_{\mathbf{k}}\,c^{\dagger}_{\mathbf{k}}c^{\dagger}_{-\mathbf{k}} + \mathrm{h.c} ) %\notag %\\
%{}& \hspace{1.0cm}
- \sum_{\mathbf k} (\Pi^{\mathbf{Q}}_{\mathbf k}\,c^{\dagger}_{\mathbf{k}}
c^{\dagger}_{-(\mathbf{k}+\mathbf{Q})}+ \mathrm{h.c} )$. %-
%\sum_{\mathbf{k}}\bigl( W_{\mathbf{k}}\,c^{\dagger}_{\mathbf{k}}c_{\mathbf{k}+\mathbf{Q}} +
%\mathrm{h.c} \bigr)
%\end{align}
%\noindent
The first term in describes a tight binding dispersion
which generically can be written as a sum of particle-hole symmetric terms
and particle-hole asymmetric terms:
%that represent the
%deviations from nesting:
$\xi_{\bf k}=\gamma_{\bf k}+\delta_{\bf k}$.
When $\delta_{\mathbf{k}}=0$ there is particle-hole symmetry or perfect nesting while finite values of $\delta_{\mathbf{k}}$ destroy the nesting conditions.
The second term $\Delta^{\mathbf{0}}_{\mathbf{k}}= \sum_{\mathbf{k'}} V^{\mathbf{0}}_{\mathbf{k},\mathbf{k'}}
\left< c_{-\mathbf{k'}} c_{\mathbf{k'}} \right> $ represents unconventional SC with zero pair momentum,
and the last term  $\Pi^{\mathbf{Q}}_{\mathbf{k}}=\sum_{\mathbf{k'}} V^{\mathbf{Q}}_{\mathbf{k},\mathbf{k'}}
\left< c_{-(\mathbf{k'}+\mathbf{Q})} c_{\mathbf{k'}} \right> $ is the TCPDWSC or modulated SC state.
Although our TCPDWSC bear some resemblance with the FFLO state \cite{FFLO}
because Cooper pairs have finite total pair momentum and the
the superfluid density is inhomogeneous, they are fundamentally different.
In fact, our TCPDWSC is a spin-triplet state whereas the FFLO is a spin-singlet
trying to survive the Zeeman field. The wavevector of the superfluid modulation
in our TCPDWSC is the {\it commensurate nesting vector $\mathbf{Q}$}.
In the FFLO state instead, the wavevector of the superfluid modulation
is variable scaling with the magnitude of the magnetic field.

The effective interactions of the itinerant quasiparticles
$V^{\mathbf{0}}_{\mathbf{k},\mathbf{k'}}$, $V^{\mathbf{Q}}_{\mathbf{k},\mathbf{k'}}$
may have a purely electronic origin in the case of FM superconductors.
However, our approach is generic irrespective of the microscopic
origin of the effective interactions, and the validity of our findings is generic as well.
In the case of heterostructures, we assume within our approach that
the effective potentials incorporate the proximity effects as well.
Naturally, we would expect in that case
a real space dependence of the potentials, that we neglect here. We only focus
on qualitative symmetry questions that would not be affected by a smooth space dependence.
In fact, the modulation of the superfluid density in our TCPDWSC state has a wavelength negligible
compared to the coherence length and the characteristic lengths of the heterostructure.
We therefore expect our qualitative findings to hold for bulk materials and for nanostructures as well.

To treat both types of SC
order parameters (OPs) in a compact
manner we introduce a Nambu-type representation using the spinors
$\Psi^{\dagger}_{\bf{k}}=\bigl(c^{\dagger}_{\bf{k}},
c_{-\bf{k}},c^{\dagger}_{\bf{k}+\bf{Q}},c_{-\bf{k}-\bf{Q}}\bigr) $.
and we use the basis provided by the tensor products
$\widehat{\rho}_{i}=\bigl(\widehat{\sigma}_{i}\otimes \hat{1}_{2})~
\mbox{and}~ \widehat{\sigma}_{i}=\bigl(\hat{1}_{2} \otimes
\widehat{\sigma}_{i})$, where \(\widehat{\sigma}_{i}\) with \(i=1,2,3\) are the usual 2x2 Pauli matrices
and \( \hat{1}_{2} \) the unit 2x2 matrix.
The absence of spin index in the hamiltonian
affects the symmetry classification of the acceptable
\emph{triplet} SC states.
%that we report below.
for which we produced a
systematic phase map.
 In fact, the OPs are normally classified by their behavior under inversion $(\hat{I})$ $\mathbf{k} \rightarrow -\mathbf{k}$, translation
$(\hat{t}_{\bf {Q}})$ $\mathbf{k} \rightarrow \mathbf{k}+\mathbf{Q}$ and time reversal $(\hat{T})$.

Instead of the latter we may use complex conjugation $(\hat{K})$ which
is related to time reversal via the relations $\hat{T} \equiv -\hat{K}(\Delta^{\mathbf{0}}_{\mathbf{k}})$ and
$\hat{T} \equiv \hat{I} \hat{K}(\Delta^{\mathbf{Q}}_{\mathbf{k}})$.
Since the spins are frozen, the homogeneous ($\mathbf{q}=0$) SC pair states may only have
odd parity: $\Delta^{\mathbf{0}}_{-\mathbf{k}} = -\Delta^{\mathbf{0}}_{\mathbf{k}}$.
Under translation we have both signs $\Delta^{\mathbf{0}}_{\mathbf{k+Q}}=
\pm\Delta^{\mathbf{0}}_{\mathbf{k}}$ and under $\hat{T}$ we get
$\hat{T} \Delta^{\mathbf{0}}_{\mathbf{k}}= - \Delta^{\mathbf{0}\,*}_{\mathbf{k}}$.
TCPDWSC states may have both parities $\Pi^{\mathbf{Q}}_{-\mathbf{k}}= \pm \Pi^{\mathbf{Q}}_{\mathbf{k}}$ and
both signs under translation since
$\Pi^{\mathbf{Q}}_{\mathbf{k+Q}}= -\Pi^{\mathbf{Q}}_{-\mathbf{k}} = \mp \Pi^{\mathbf{Q}}_{\mathbf{k}}$. Time reversal
demands that $\hat{T}\Pi^{\mathbf{Q}}_{\mathbf{k}}=\Pi^{\mathbf{Q}\,*}_{-\mathbf{k}}$ implying the relation $\hat{T}=\hat{I} \hat{K}$ for the
TCPDWSC states.
The break of time reversal allows finally four possible SC
OPs, two homogeneous SC states and two TCPDWSC states:
$ \Delta^{\mathbf{0}I--}_{\mathbf{k}}, \hskip 0.3cm \Delta^{\mathbf{0}I-+}_{\mathbf{k}},
\hskip 0.3cm \Pi^{\mathbf{Q}I-+}_{\mathbf{k}}, \hskip 0.3cm
\Pi^{\mathbf{Q}R+-}_{\mathbf{k}}$.
Here the first index $\mathbf{q}=\mathbf{0}$ or $\mathbf{q}=\mathbf{Q}$ indicates the {\it total
momentum of the pair} (or the characteristic wavevector of the superfluid density), the second index $R$ or $I$ indicates whether
the OP is real or imaginary, the third index $\pm$
indicates parity under inversion $\hat{I}$ and the last index denotes gap
symmetry under $\hat{t}_{\mathbf{Q}}$.
The symmetry properties of the OPs under inversion \(\hat{I} \)
and translation \(\hat{t}_{\mathbf{Q}} \) imply a specific structure in \(\mathbf{k}\)-space.
Every OP $M_{\mathbf{k}}$ is written in the form $M_{\mathbf{k}}=M f_{\mathbf{k}}$
where the {\it form factors $f_{\mathbf{k}}$} belong to the different irreducible
representations of the point group.

According to the above symmetry classification there exist
four possible pairs of competing homogeneous and modulated SC states.
Using our formalism we calculate Green's functions and from them
self-consistent systems of coupled gap equations for each case.
The pairs $\Delta^{\mathbf{0}I--}_{\mathbf{k}}$ with $\Pi^{\mathbf{Q}R+-}_{\mathbf{k}}$ and
$ \Delta^{\mathbf{0}I--}_{\mathbf{k}}$ with $\Pi^{\mathbf{Q}I-+}_{\mathbf{k}} $
obey the system of coupled equations:
$\Delta_{\mathbf{k}} = \sum_{\mathbf{k'}} V^{\Delta}_{\mathbf{k},\mathbf{k'}} \Delta_{\mathbf{k'}}
\sum_{\pm}
\frac{1}{4E_{\pm}(\mathbf{k'})} \tanh ({ E_{\pm}(\mathbf{k'})\over 2T })$
and
$\Pi_{\mathbf{k}} = \sum_{\mathbf{k}'} V^{\Pi}_{\mathbf{k}, \mathbf{k'}} \Pi_{\mathbf{k}'}
\sum_{\pm}
{A_{\mathbf{k}'} \pm \gamma_{\mathbf{k}'}
\over 4 E_{\pm}({\mathbf{k}'})A_{\mathbf{k}'}}
\tanh ( {E_{\pm} ({\mathbf{k}'})\over 2T} )$%\\% \nonumber\\
where $A_{\mathbf{k}} \equiv %\sqrt{
[\delta_{\mathbf{k}}^2 + \Pi_{\mathbf{k}}^2]^{1/2}$ and the quasiparticle energies are
$E_{\pm}(\mathbf{k}) = %\sqrt{
[(\sqrt{\delta^2_{\mathbf{k}}+ \Pi^2_{\mathbf{k}}} \pm \gamma_{\mathbf{k}})^{2}
+\Delta^2_{\mathbf{k}}]^{1/2}$.
The remaining two cases, competition of $ \Delta^{\mathbf{0}I-+}_{\mathbf{k}}$ with $\Pi^{\mathbf{Q}R+-}_{\mathbf{k}} $ and
$ \Delta^{\mathbf{0}I-+}_{\mathbf{k}}$ with $\Pi^{\mathbf{Q}I-+}_{\mathbf{k}} $, obey the following equations:
$\Delta_{\mathbf{k}} = \sum_{\mathbf{k'}} V^{\Delta}_{\mathbf{k},\mathbf{k'}} \Delta_{\mathbf{k'}}
\sum_{\pm}
\frac{B_{\mathbf{k}'}\pm \Pi^{2}_{\mathbf{k'}}}{4E_{\pm}(\mathbf{k'})B_{\mathbf{k}'}}
\tanh ({ E_{\pm}(\mathbf{k'})\over 2T })$ %\\ %\nonumber \\
and $\Pi_{\mathbf{k}} = \sum_{\mathbf{k}'} V_{\mathbf{k}, \mathbf{k'}}^{\Pi} \Pi_{\mathbf{k}'}
\sum_{\pm}
\frac{B_{\mathbf{k}'}\pm \gamma^{2}_{\mathbf{k'}}\pm \Delta^{2}_{\mathbf{k'}}}{4E_{\pm}(\mathbf{k'})B_{\mathbf{k}'}} \tanh ({ E_{\pm}(\mathbf{k'})\over 2T })$
where $B_{\mathbf{k}} \equiv %\sqrt{
[\gamma^{2}_{\mathbf{k}}A_{\mathbf{k}}^{2} + \Delta^{2}_{\mathbf{k}}\Pi^{2}_{\mathbf{k}}]^{1/2}$ and
the dispersions are
$E_{\pm}({\bf k}) = %\sqrt{
\bigl[\Delta_{\mathbf{k}}^2 \delta_{\mathbf{k}}^2 A_{\mathbf{k}}^{-2}
+ ( A_{\mathbf{k}} \pm \sqrt{\gamma_{\mathbf{k}}^2 +  \Delta_{\mathbf{k}}^2
\Pi_{\mathbf{k}}^2 A_{\mathbf{k}}^{-2}}\,)^2\bigr]^{1/2}$.

The effective potentials $V^{\Delta}_{\mathbf{k},\mathbf{k'}}, V^{\Pi}_{\mathbf{k},\mathbf{k'}}$ have the form
$V_{\mathbf{k},\mathbf{k'}}= V f_{\mathbf{k}}f_{\mathbf{k}'} $ (separable potentials).
We have solved self consistently the above systems of equations
on a square lattice with $\gamma_{\mathbf{k}}= -t_1 (\cos k_x + \cos k_y)$ and
$\delta_{\mathbf{k}}=- t_2\cos k_x \cos k_y$ and $\mathbf{Q}=(\pi,\pi)$.
The choice of a {\it tetragonal} dispersion is motivated by the fact that CrO$_2$ as well as strongly FM superconductors like UGe$_2$ and URhGe exhibit all a tetragonal structure, however, our qualitative findings are generic.
The corresponding form factors belong to irreducible representations of the tetragonal group D$_{4h}$. Specifically:  $\Delta^{\mathbf{0}I--}_{\mathbf{k}} \sim \sin k_{x} + \sin k_{y}$ (s-wave),
$\Delta^{\mathbf{0}I-+}_{\mathbf{k}}, \Pi^{\mathbf{Q}I-+}_{\mathbf{k}} \sim sin (k_{x}+k_{y})$ (p-wave) and
$\Pi^{\mathbf{Q}R+-}_{\mathbf{k}} \sim \cos k_{x} - \cos k_{y}$ (d-wave).
For every competing pair we have performed a large number of self-consistent calculations
varying pairing potentials in the two channels, temperatures
and ratios $t_{2}/t_{1}$.

%%%%%%%%%%%%%%%%%%%%%%%%%%%%%%%%%%%%%%%%%%%%%%%%%%%%%%%%%%%%%%%%%%%%%%%%
\begin{figure}[!h]
\includegraphics[width=7.5cm,height=10.0cm,angle=0]{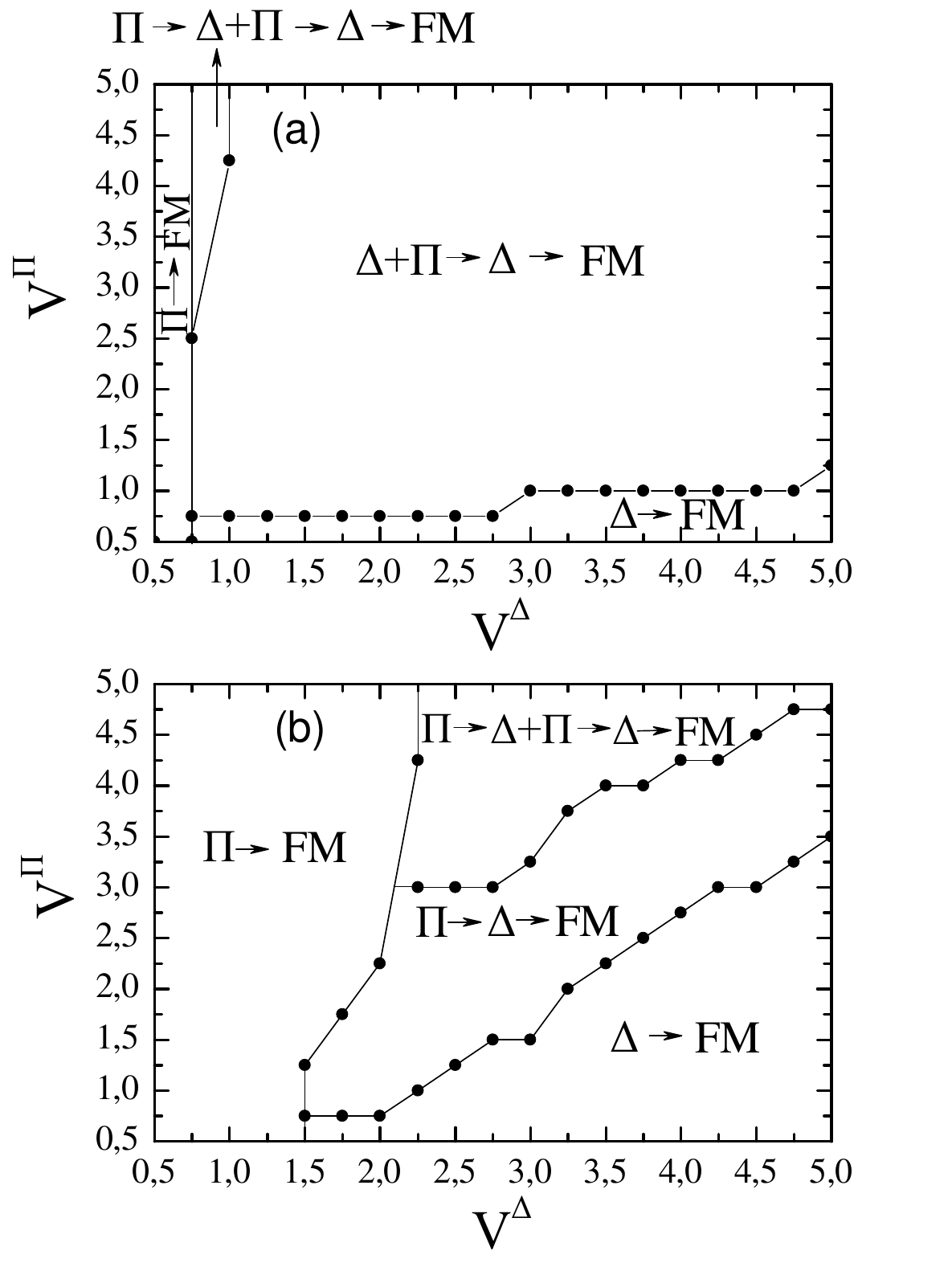}
\caption{Maps of the dependence of phase sequences on the effective interactions
$V^{\Delta}$ and $V^{\Pi}$ for low temperature. Arrows indicate
the cascade of phases obtained when $t_2/t_1$ grows starting from zero. The black dots
separate regions of different phase {\it sequences} under growing $t_2/t_1$.
All phases coexist with ferromagnetism (FM). The phases
indicated as FM only FM is present. Panel (a) depicts the interplay of $\Pi^{\mathbf{Q}R+-}$ with $\Delta^{\mathbf{0}I--}$
whereas panel (b) that of $\Pi^{\mathbf{Q}R+-}$ with $\Delta^{\mathbf{0}I-+}$.
The potentials are in units of $t_1$.} \label{fig:maps}
\end{figure}
%%%%%%%%%%%%%%%%%%%%%%%%%%%%%%%%%%%%%%%%%%%%%%%%%%%%%%%%%%%%%%%%%%%%%%%%%

The first important result is that the TCPDWSC $\Pi^{\mathbf{Q}I-+}_{\mathbf{k}}$ OP
can never survive. %when it competes with any of the two zero pair momentum SC states.
Specifically, the $\Pi^{\mathbf{Q}I-+}_{\mathbf{k}}$ gap is zero regardless of the values of the pairing potentials and the
particle-hole asymmetry $t_{2}/t_{1}$ term. We conclude that
\emph{although the state $\Pi^{\mathbf{Q}I-+}_{\mathbf{k}}$ is allowed by symmetry, it is never realized}.
Therefore, we only report results about the relevant competition of the remaining TCPDWSC OP $\Pi^{\mathbf{Q}R+-}_{\mathbf{k}} $ with both zero momentum SC states.

The phase sequences as $t_{2}/t_{1}$ grows starting from zero
and for various values of the pairing potentials for the
competition $\Pi^{\mathbf{Q}R+-}_{\mathbf{k}} $ with $\Delta^{\mathbf{0}I--}_{\mathbf{k}}$ and
$\Pi^{\mathbf{Q}R+-}_{\mathbf{k}} $ with $\Delta^{\mathbf{0}I-+}_{\mathbf{k}}$ are shown in
the respective panels of Fig. \ref{fig:maps}.
Arrows in Fig. \ref{fig:maps} indicate the cascade of phases
observed when the ratio $t_{2}/t_{1}$ grows starting from zero
at each region of the map. The variation of $t_{2}/t_{1}$
may simulate various effects such that chemical doping, or stress effects
as well as proximity effects.
Since we consider a spin-polarized background, all states reported coexist with FM,
and the transitions to the FM state reported at high values of $t_{2}/t_{1}$ has the meaning
of a transition to a state that is only ferromagnetic with
no SC OP present.

Both cases share the characteristic feature that the TCPDWSC state $\Pi^{\mathbf{Q}R+-}_{\mathbf{k}}$
is finite in the largest part of the maps of the pairing potentials.
Thus, since we do not limit to a specific microscopic model that could
correspond to a specific value for the
pairing potentials, the existence of the modulated TCPDWSC phase can be considered
generically plausible.
The interplay of $\Pi^{\mathbf{Q}R+-}_{\mathbf{k}}$ with $\Delta^{\mathbf{0}I--}_{\mathbf{k}}$ favors the \emph{coexistence} of both
$(\mathbf{q}=\mathbf{0} \; \text{and} \; \mathbf{q}=\mathbf{Q})$ SC states at low-T
over a wide range of values of the pairing potentials (Fig. \ref{fig:maps}a).
The transition from a coexistence state to a homogeneous ($\mathbf{q}=\mathbf{0}$) SC state
 as $t_2/t_1$ grows is always continuous (\emph{second order}) and dominates the $V^{\Delta},V^{\Pi}$ parameter space.

The low temperature regime is different in the interplay of $\Pi^{\mathbf{Q}R+-}_{\mathbf{k}}$ with $\Delta^{\mathbf{0}I-+}_{\mathbf{k}}$.
Coexistence of the two SC states is allowed again but now is restricted to a small portion of the $V^{\Delta},V^{\Pi}$
map (Fig. \ref{fig:maps}b).
The most interesting feature is now the \emph{the domination
of the TCPDWSC (modulated SC) state for the smaller values of $t_2/t_1$.} Thus, in this case the formation of
the $\Pi^{\mathbf{Q}R+-}$ TCPDWSC state is favored.
As particle hole asymmetry grows ($t_{2}/t_{1}$ grows) we may have transitions from
TCPDWSC to a state of coexistence or to a homogeneous SC state.

\begin{figure}[!h]
\centerline{\includegraphics[width=8cm,height=6cm]{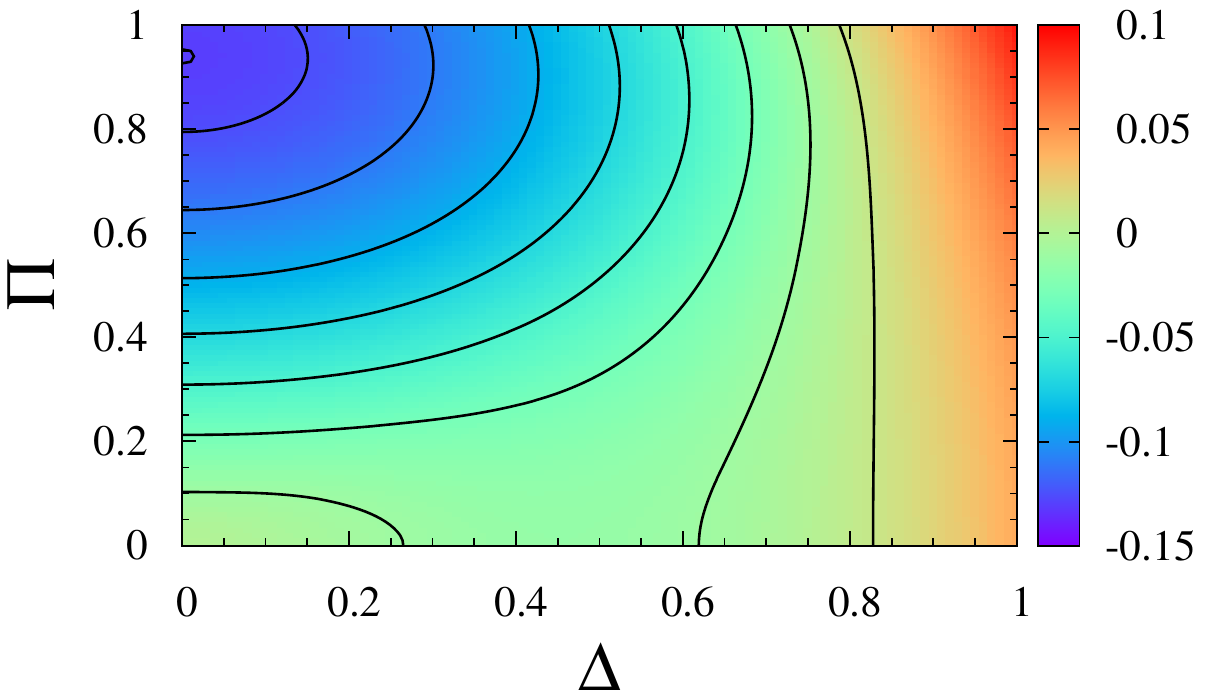}}
%\centerline{\includegraphics[width=5cm,height=3cm]{free_new.pdf}}
\caption {{\footnotesize (Color online)
Contour plot of the condensation free energy
$\Delta F$ as a function of
the OPs $\Pi^{\mathbf{Q}R+-}_{\mathbf{k}}$ and $\Delta^{\mathbf{0}I-+}_{\mathbf{k}}$ at low-T for $t_{2}/t_{1}=0$ and $V^{\Delta}=V^{\Pi}=3t_1$.
The lowest free energy is situated at the point $(\Delta,\Pi)=(0,0.94t_{1})$
where only $\Pi^{\mathbf{Q}R+-}_{\mathbf{k}}$
is finite despite the fact that it exhibits gapless SC.}} \label{fig:free-energy}
\end{figure}

The stability of the solutions of the self consistent-equations has been verified by free-energy calculations
as well. The free-energy difference $\Delta F$
between the normal and the condensed state is given by:
%\vspace{-0.1cm}
%\begin{align}
%\label{free-energy}
$\Delta F = \frac{\Delta^{2}}{V^{\Delta}} + \frac{\Pi^{2}}{V^{\Pi}}
- \frac{1}{2\beta} \sum_{\mathbf{k}}\sum_{j=\pm, i=\pm}  \mathrm{ln} (\frac{1+e^{-j\beta E_{i}(\mathbf{k})}}{1+e^{-j\beta \epsilon_{i}(\mathbf{k})}})$
%\end{align}
%
%\noindent
where
%$\Delta$,$\Pi$ are the gaps, $V^{\Delta}$,$V^{\Pi}$ the magnitudes of their
%respective potentials,
$E_{\pm}({\bf k})$ the energy dispersions for each competing pair and
$\epsilon_{\pm}({\bf k})$ the energy dispersions obtained when both gaps are zero.
The safest way to ensure that the solutions of the coupled gap equations correspond to
the minimum of the free-energy difference is to vary $\Delta F$ with respect to
the magnitudes of the gaps and verify that
$\Delta F$ attains its minimum for these values.
We report in Fig. \ref{fig:free-energy} the variations
of the free-energy difference with
$\Delta^{\mathbf{0}I-+}_{\mathbf{k}}$ and $\Pi^{\mathbf{Q}R+-}_{\mathbf{k}}$
at low-T for $t_{2}/t_{1}=0$
%in case $\Pi^{\mathbf{Q}R+-}_{\mathbf{k}}$ competes with $\Delta^{\mathbf{0}I-+}_{\mathbf{k}}$
and for  $V^{\Delta}=V^{\Pi}=3$
to illustrate the dominance of the TCPDWSC state. These values of the pairing potentials
correspond to the cascade of transitions $\Pi \rightarrow \Delta \rightarrow FM$
when $t_2/t_1$ grows (cf. Fig. \ref{fig:maps}b).
%\vspace{-0.5cm}
It is clear that $\Delta F$ attains its minimum value
for $(\Delta,\Pi)=(0,0.94t_{1})$, thus the ground state consists solely
of theTCPDWSC phase. \\

%%%%%%%%%%%%%%%%%%%%%%%%%%%%%%%%%%%%%%%%%%%%%%%%%%%%%%%%%%%%%%%%%%%%%%%%
\begin{figure}[!h]
\centerline{\includegraphics[width=7cm,height=8cm]{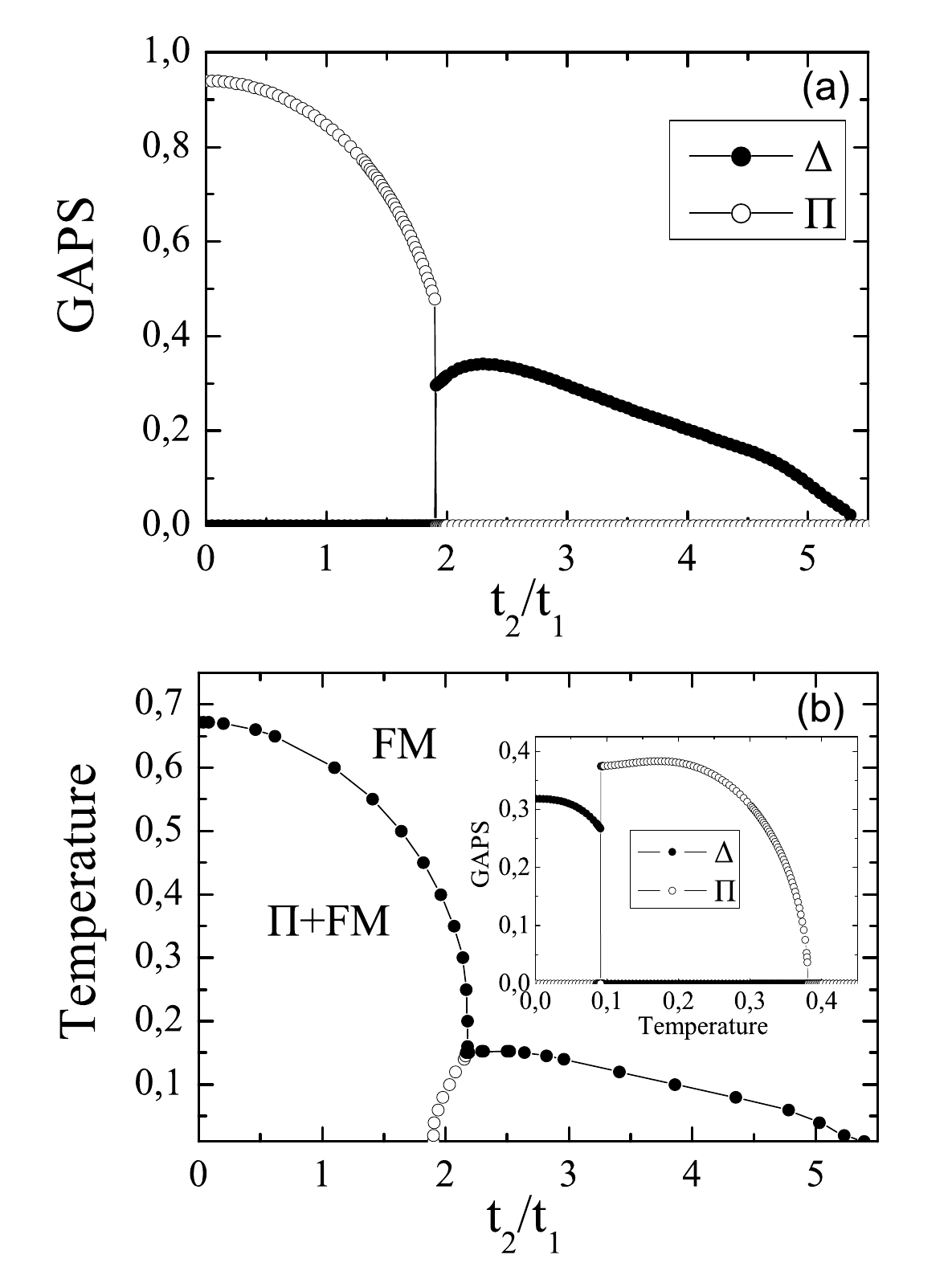}}
%\centerline{\includegraphics[width=5cm,height=5.5cm]{Figu2.pdf}}
\caption {{\footnotesize (a) Dependence of homogeneous $\Delta^{\mathbf{0}I-+}_{\mathbf{k}}$
and modulated $\Pi^{\mathbf{Q}R+-}_{\mathbf{k}}$ SC gaps
on \(t_{2}/t_{1}\) at low-T. (b) $t_{2}/t_{1}$-temperature phase diagram.
Closed symbols mark 2nd order and open symbols 1st order transitions.
A first order transition, for $t_{2}/t_{1}=2$, \emph{within the SC phase} from the TCPDWSC to homogeneous SC,
is possible with decreasing temperature (inset).
The values of the pairing potentials are  $V^{\Delta}=V^{\Pi}=3t_1$.  }} \label{ZCI-+ZBR+-:phase}
\end{figure}
%%%%%%%%%%%%%%%%%%%%%%%%%%%%%%%%%%%%%%%%%%%%%%%%%%%%%%%%%%%%%%%%%%%%%%%%
We report in Fig. \ref{ZCI-+ZBR+-:phase} the dependence of the OPs on $t_{2}/t_{1}$
at low-T (Fig. \ref{ZCI-+ZBR+-:phase}a) and the phase diagram (Fig. \ref{ZCI-+ZBR+-:phase}b) obtained
by the coupled-gap equations. %(\ref{zones2ZCgap})-(\ref{zones2ZBgap}).
We stress that at low-T for $t_{2}/t_{1}=0$ the values of the gaps
are in full agreement with the $\Delta F$ minimum requirement, i.e $(\Delta,\Pi)=(0,0.94t_{1})$.
The $t_{2}/t_{1}$ transition from the modulated to the homogeneous SC state is \emph{first order},
and we note that the TCPDWSC gap is significantly larger than the homogeneous SC gap despite the fact
 that the pairing potentials have the same magnitude (Fig. \ref{ZCI-+ZBR+-:phase}a).
The phase diagram shows that the transition $\Pi \rightarrow \Delta$ with $t_{2}/t_{1}$ is not limited to low-T.
The modulated SC phase extends to higher temperatures (Fig. \ref{ZCI-+ZBR+-:phase}b) than the homogeneous SC phase.
The boundary separating the two SC states remains \emph{first order} and ends at a \emph{tricritical} point.
Decreasing the temperature moves the boundary to lower $t_{2}/t_{1}$-values.
This allows a \emph{first order}  transition with respect to temperature \emph{within the superconducting phase} from the TCPDWSC to the homogeneous SC state. An example of such a transition realized for $t_{2}/t_{1}=2$ is shown in the inset of Fig. \ref{ZCI-+ZBR+-:phase}b.

\begin{figure}[!h]
\includegraphics[width=4cm,height=4cm]{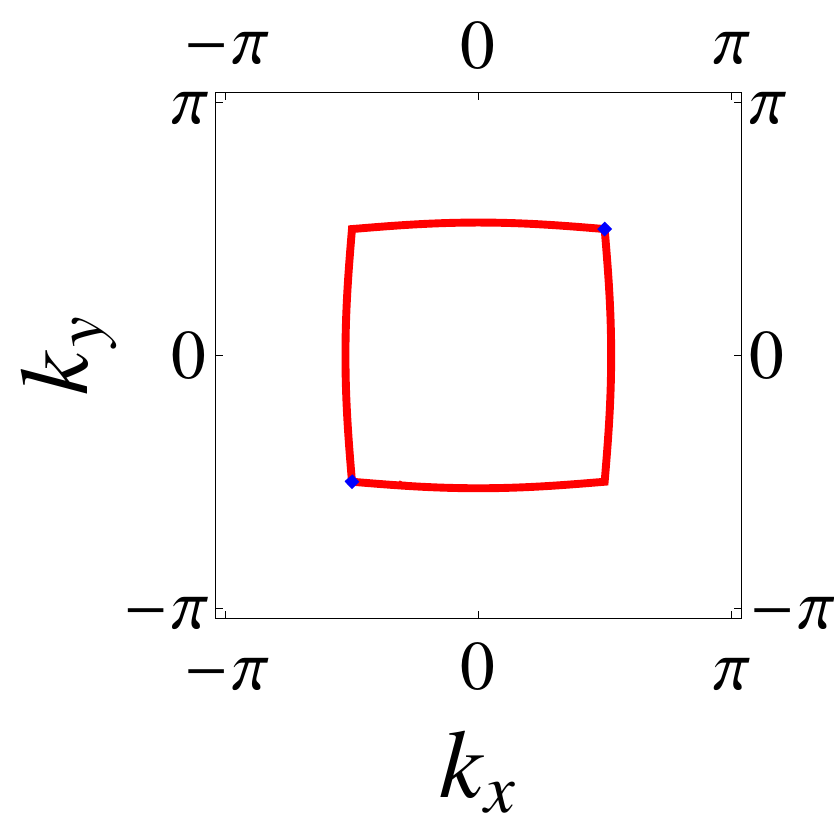}
\includegraphics[width=4.5cm,height=4cm]{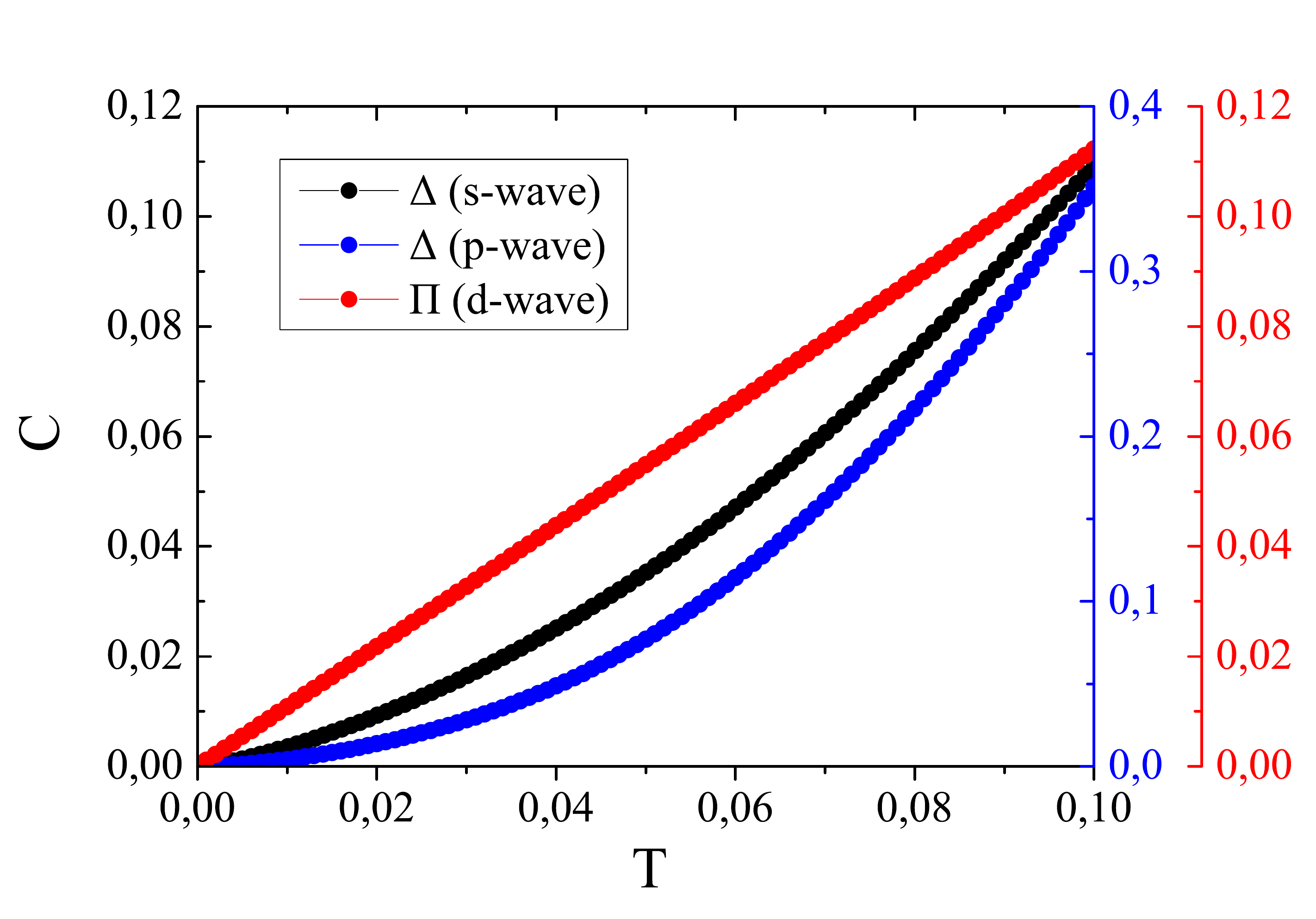}
\caption{{\footnotesize (Color online) Fermi surface (left) and specific heat at low-T  (right) in the $\Pi^{\mathbf{Q}R+-}_{\mathbf{k}}$ state (red) for $t_{2}/t_{1}=1.0$ and the
$\Delta^{\mathbf{0}I-+}_{\mathbf{k}}$ state  (blue) for $t_{2}/t_{1}=2.5$.
The extended FS in the TCPDWSC state causes the linear behavior of the specific heat, whereas the
polynomial dependence in the $\Delta^{\mathbf{0}I-+}_{\mathbf{k}}$ state is a direct consequence
of the presence of Fermi points instead of FS. The pairing potentials are  $V^{\Delta}=V^{\Pi}=3t_1$.} } \label{fig:FermiCv}
\end{figure}

A question that naturally arises is how the exotic TCPDWSC state $\Pi^{\mathbf{Q}R+-}_{\mathbf{k}}$
can be identified experimentally.
Quite remarkably, specific heat measurements at low-T may be sufficient.
Specifically, isolated TCPDWSC states exhibit an extended FS whereas the FS is limited
to two Fermi points
in the coexistence phase $\Delta + \Pi$ .
Therefore a polynomial behavior of the specific heat at low-T is a signature of the coexistence phase.
As particle-hole asymmetry $t_{2}/t_{1}$ grows for example with gate voltage, only the modulated SC state
$\Pi^{\mathbf{Q}R+-}_{\mathbf{k}}$ continues to exhibit extended FS whereas the zero momentum SC states as well as the coexistence phase present limited FS consisting of \emph{isolated Fermi points}.
We note that the extended FS is also a feature of the spin-singlet $\eta$-pairing \cite{Fradkin1}.%}

Consequently the TCPDWSC state $\Pi^{\mathbf{Q}R+-}_{\mathbf{k}}$ is the sole SC state
that exhibits a linear low-T behavior of the specific heat and this is robust since it
holds even for finite
values of $t_{2}/t_{1}$.
This is illustrated in Fig. \ref{fig:FermiCv} where the Fermi surface and the specific heat
for $t_{2}/t_{1}=1.0$  in the TCPDWSC phase (red) and
$t_{2}/t_{1}=2.5$  in the homogeneous SC phase (blue) of Fig. \ref{ZCI-+ZBR+-:phase} are reported.
We observe that in the $\Pi^{\mathbf{Q}R+-}_{\mathbf{k}}$ phase the
Fermi surface is extended imposing the linear behavior of the specific heat at low-T.
On the other hand, in the
$\Delta^{\mathbf{0}I-+}_{\mathbf{k}}$ phase we only have two Fermi points and the specific heat at low-T exhibits a polynomial behavior. This is also the case for the other homogeneous SC state $\Delta^{\mathbf{0}I--}_{\mathbf{k}}$ as well as for the coexistence phase $\Delta+\Pi$.
Therefore \emph{linear low-T specific heat in the SC state identifies the triplet TCPDWSC state}.

\begin{figure}[!h]
\includegraphics[width=4.cm,height=4cm]{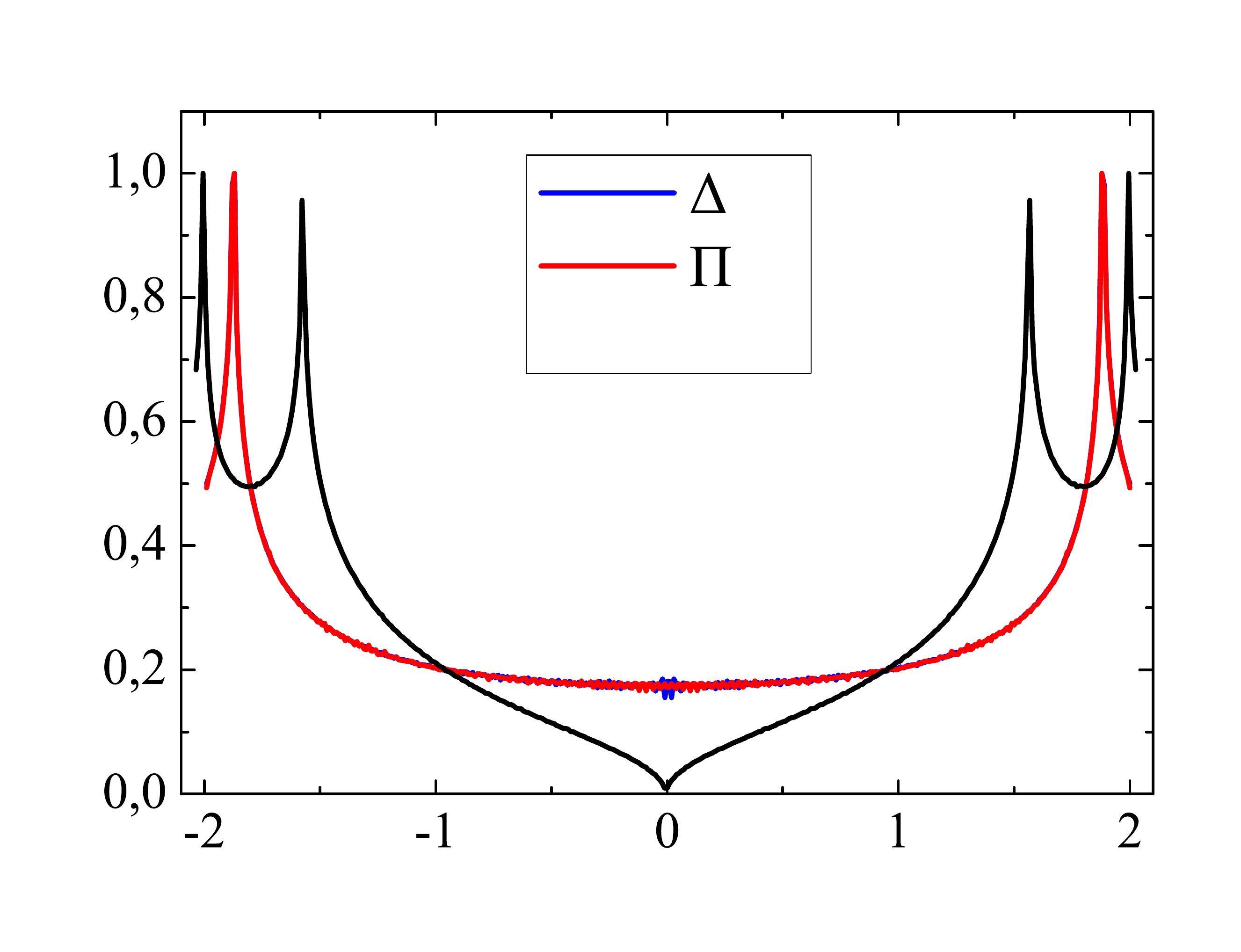}
\includegraphics[width=4.cm,height=4cm]{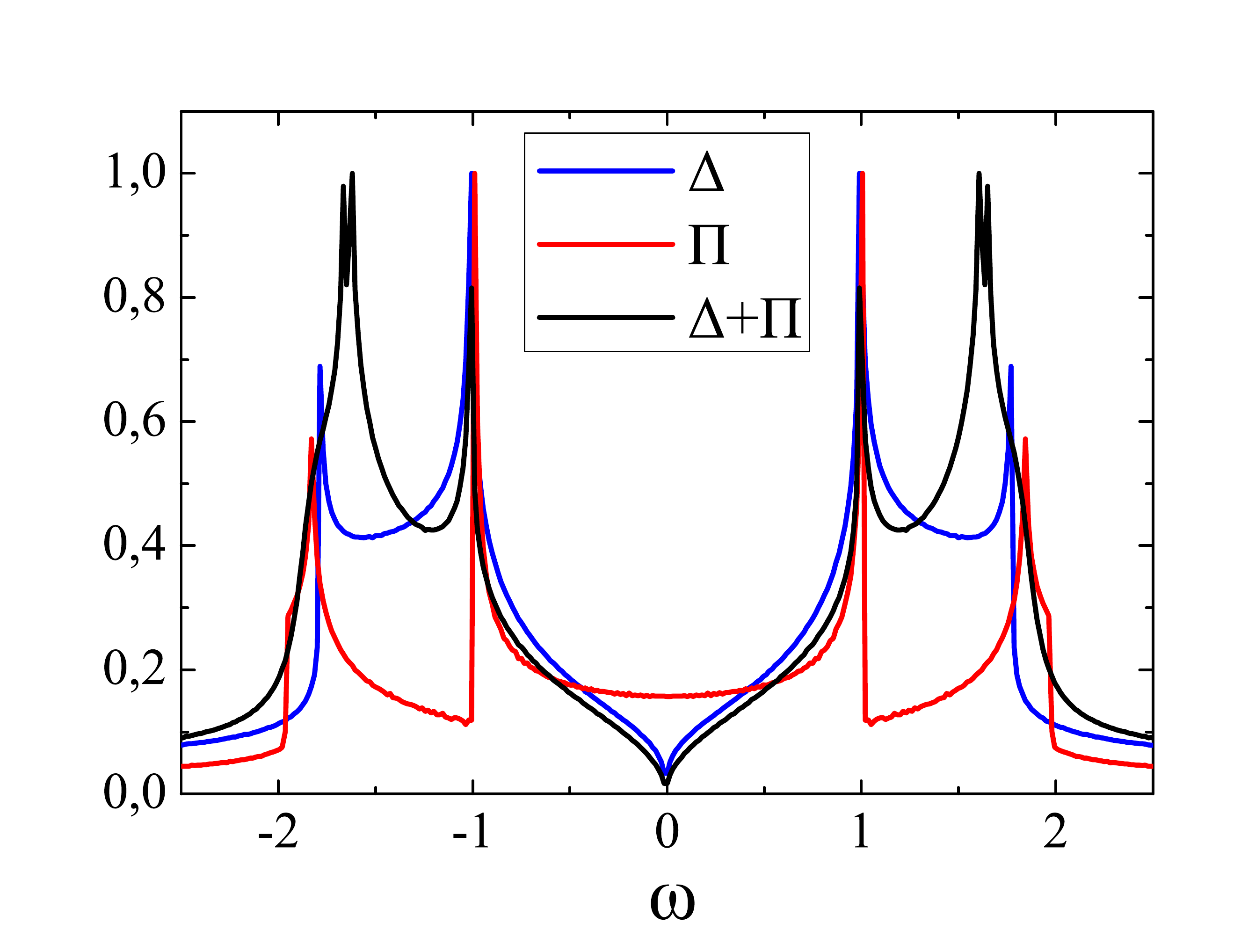}
\caption{{\footnotesize (Color online)
DOS for $t_{2}/t_{1}=0$ (left) and $t_{2}/t_{1}=1$ (right) at low-T in the
$\Delta^{\mathbf{0}I--}_{\mathbf{k}}$ (blue), the $\Pi^{\mathbf{Q}R+-}_{\mathbf{k}}$ (red) and the
coexistence phase $\Delta + \Pi$. The pairing potentials are equal $V^{\Delta,\Pi}=3t_1$.
} } \label{fig:DOS}
\end{figure}
The difference in the FS is reflected in the behavior of the electronic density of states (DOS)
N($\omega$)
accessible by tunneling.
In our spinor formalism: N($\omega$)= $-\frac{1}{\pi} \mathsf{Im} \sum_{\mathbf{k}} \mathsf{Tr} \{ \mathcal{G}(\mathbf{k},\mathrm{i}\omega_{n} \rightarrow \omega + \mathrm{i} n)\}$.
Performing the analytical continuation it can be shown to take the form:
%\begin{align}
%\label{DOS}
$N(\omega) = \sum_{\mathbf{k}} \{ \delta\bigl(\omega + E_{\pm}(\mathbf{k})\bigr) +  \delta\bigl(\omega - E_{\pm}(\mathbf{k})\bigr) \}$.
As an example we present in Fig. \ref{fig:DOS} the DOS in the $\Pi^{\mathbf{Q}R+-}_{\mathbf{k}}$, the $\Delta^{\mathbf{0}I--}_{\mathbf{k}}$ state  and the coexistence phase $\Delta+\Pi$ for $t_{2}/t_{1}=0$ (left) and
$t_{2}/t_{1}=1$ (right).
In each case the pairing potential is 3$t_{1}$.
The vanishing DOS at the Fermi level identifies the coexistence phase $\Delta+\Pi$ in the case of perfect nesting $t_{2}/t_{1}=0$,
whereas the finite DOS for particle-hole asymmetry $t_{2}/t_{1}\neq 0$ is a direct signature
of the TCPDWSC state.
We note that finite DOS at the Fermi level has also been reported in spin-singlet
PDW states \cite{Loder}.

\begin{figure}[!h]
\includegraphics[width=7cm,height=4.5cm]{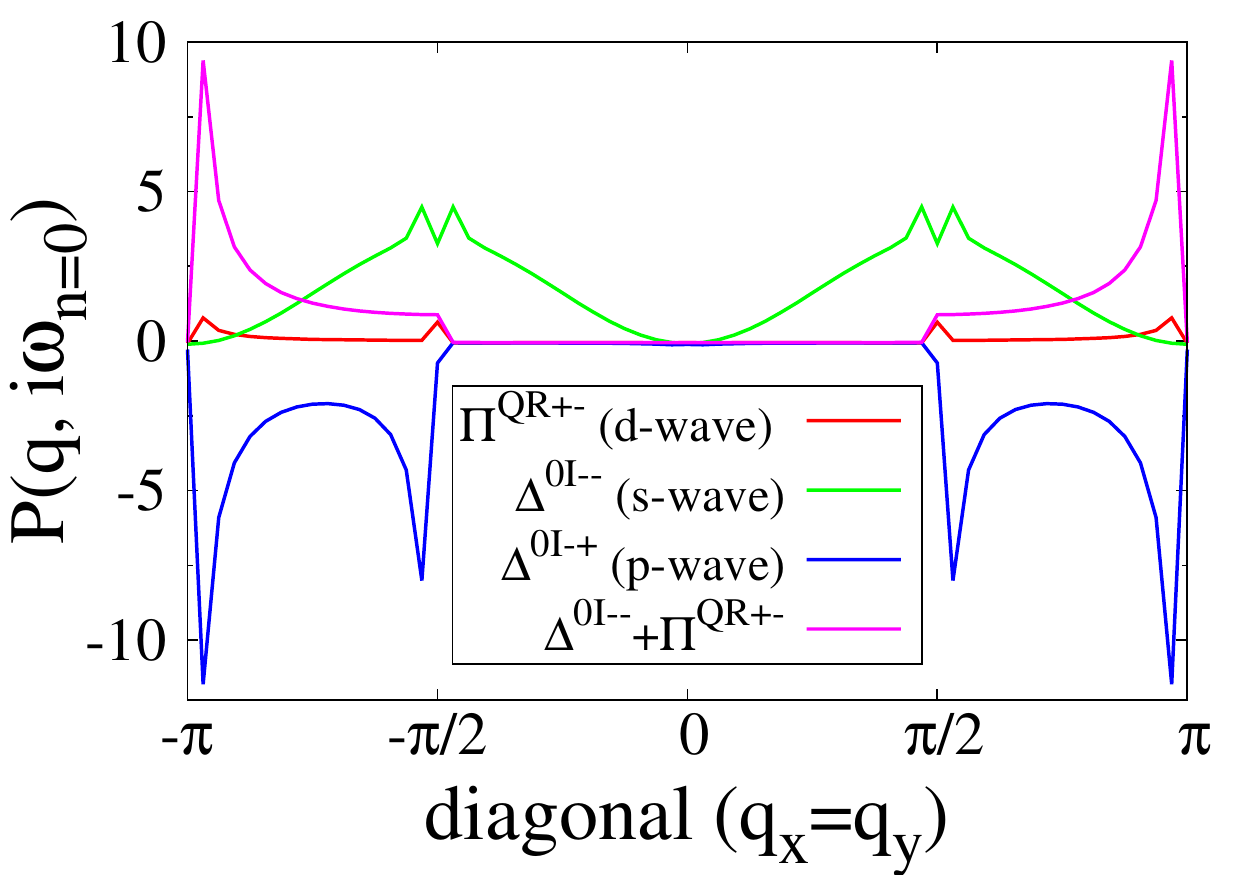}
\includegraphics[width=7cm,height=4.5cm]{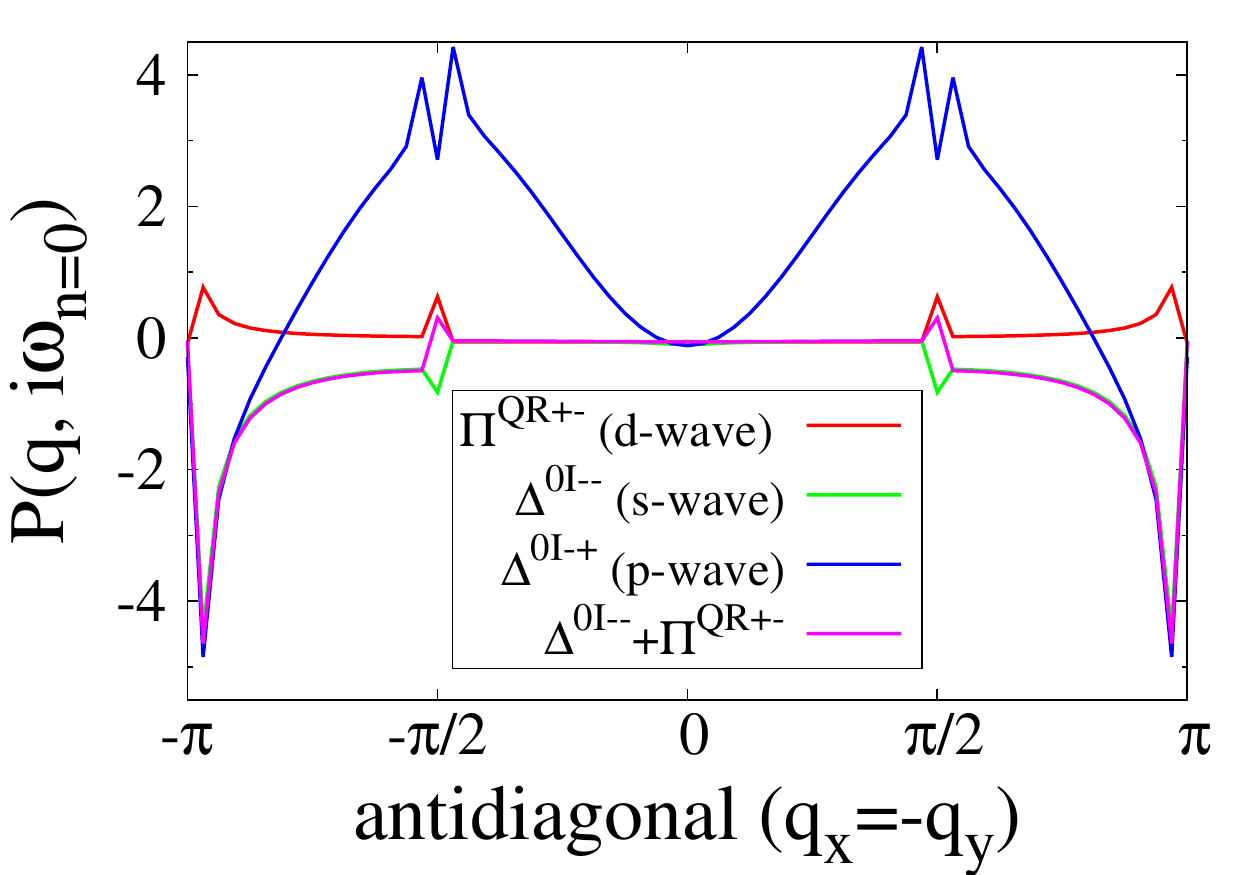}
\vspace{0.5cm}
\caption{{\footnotesize (Color online) Charge-charge correlation function $P(\mathbf{q}, \mi \omega_{n=0})$
along the diagonal (left) and the antidiagonal (right) of the FBZ in the $\Pi^{\mathbf{Q}R+-}_{\mathbf{k}}$ (red), the
$\Delta^{\mathbf{0}I--}_{\mathbf{k}}$ (green), the $\Delta^{\mathbf{0}I-+}_{\mathbf{k}}$ (blue) and the
coexistence phase $\Delta + \Pi$ (magenta) at low-T for $t_{2}/t_{1}=0$. The pairing potentials are $V^{\Delta,\Pi}=3t_1$.} } \label{fig:charge}
\end{figure}
Finally, measurements related with the charge-charge correlation function
$P(\mathbf{q}, \mi \omega_{n})$ for the
lowest Matsubara frequency $\omega_{n=0}= \pi T$ along the diagonal and the antidiagonal of
the first Brillouin zone (FBZ) may
provide the ultimate experimental strategy to distinguish the different
accessible triplet SC states.

We illustrate this in Fig. \ref{fig:charge} where $P(\mathbf{q}, \mi \omega_{n=0})$ is reported along the diagonal (left) and the antidiagonal (right) of the FBZ in the $\Pi^{\mathbf{Q}R+-}_{\mathbf{k}}$ (red), the $\Delta^{\mathbf{0}I--}_{\mathbf{k}}$ (green), the $\Delta^{\mathbf{0}I-+}_{\mathbf{k}}$ (blue) and the
coexistence phase $\Delta + \Pi$ (magenta).
The unique feature of the TCPDWSC state $\Pi^{\mathbf{Q}R+-}_{\mathbf{k}}$ is that it is the sole state for which the charge-charge correlation function is \emph{the same in both directions of the FBZ.}
In case of the homogeneous $\Delta^{\mathbf{0}I--}_{\mathbf{k}}$ state measuring
the correlation function along the diagonal direction reveals a double-peak structure around the points $\pm(\pi/2,\pi/2)$ and a decrease around the center of the FBZ whereas for all the other states it exhibits only one peak at $\pm(\pi/2,\pi/2)$ and remains practically constant
around the center of the FBZ.
Quite remarkably, the double-peak structure around the points $\pm(\pi/2,\pi/2)$ as well as the decrease around the center of the FBZ become characteristic features of the $\Delta^{\mathbf{0}I-+}_{\mathbf{k}}$ state along
the antidiagonal direction.
For the coexistence $\Delta + \Pi$ state,
in the diagonal direction it is the sole state that exhibits peaks only at the edges of the BZ, while in the antidiagonal direction there are again peaks at the edges of the BZ but they are on the negative side and two new smaller peaks at $\pm(\pi/2,\pi/2)$ that are on the positive side.

In summary, within a generic microscopic mean field theory we explored systematically
the interplay of all possible triplet SC states in an effectively spinless system.
We find that the inhomogeneous TCPDWSC state $\Pi^{\mathbf{Q}I-+}_{\mathbf{k}}$
having the p-wave symmetry can never survive
the two allowed by symmetry homogeneous triplet SC states.
However, the other
TCPDWSC $\Pi^{\mathbf{Q}R+-}_{\mathbf{k}}$ having
d-wave symmetry may \emph{either appear alone or coexist with the homogeneous SC OPs}
driving the phenomenon of gapless SC over a wide parameter range.
Our findings are universally applicable to any strongly ferromagnetic
system that develops superconductivity including devices designed to host
localized Majorana modes for topological quantum computation.
Geometry and the presence of one-spin triplet SC does not guarantee
the relevance of a device designed to host Majorana qubits, it should
be tested against the eventual emergence of
catastrophic gapless triplet SC and we have identified
some experimental paths for such tests.

We are grateful to Alexandros Aperis, Panagiotis Kotetes
and Georgios Livanas for illuminating discussions.

\end{document}